\documentclass[conference]{IEEEtran}
\IEEEoverridecommandlockouts
\usepackage{cite}
\usepackage{amsmath,amssymb,amsfonts}
\usepackage{algorithmic}
\usepackage{graphicx}
\usepackage{textcomp}
\usepackage{xcolor}
\usepackage{booktabs}
\def\BibTeX{{\rm B\kern-.05em{\sc i\kern-.025em b}\kern-.08em
    T\kern-.1667em\lower.7ex\hbox{E}\kern-.125emX}}
\begin{document}

\title{Long Short-Term Memory Spatial Transformer Network}

\author{\IEEEauthorblockN{1\textsuperscript{st} Shiyang Feng}
\IEEEauthorblockA{\textit{College of Information Science and Technology} \\
\textit{DongHua University}\\
ShangHai, China \\
syoung\_f@163.com}
\and
\IEEEauthorblockN{2\textsuperscript{nd} Tianyue Chen}
\IEEEauthorblockA{\textit{College of Science} \\
\textit{DongHua University}\\
ShangHai, China \\
spence\_chen@126.com}
\and
\IEEEauthorblockN{3\textsuperscript{rd} Hao Sun}
\IEEEauthorblockA{\textit{College of Information Science and Technology} \\
\textit{DongHua University}\\
ShangHai, China \\
shposion@icloud.com}
}

\maketitle

\begin{abstract}
Spatial transformer network has been used in a layered form in conjunction with a convolutional network to enable the model to transform data spatially. In this paper, we propose a combined spatial transformer network (STN) and a Long Short-Term Memory network (LSTM) to classify digits in sequences formed by MINST elements. This LSTM-STN model has a top-down attention mechanism profit from LSTM layer, so that the STN layer can perform short-term independent elements for the statement in the process of spatial transformation, thus avoiding the distortion that may be caused when the entire sequence is spatially transformed. It also avoids the influence of this distortion on the subsequent classification process using convolutional neural networks and achieves a single digit error of 1.6\% compared with 2.2\% of Convolutional Neural Network with STN layer.
\end{abstract}


\section{Introduction}
The attention mechanism in computer vision has been applied in object detection \cite{walther2006modeling}\cite{viola2001rapid}\cite{xu2015show}\cite{han2018deep} \cite{lin2018focal}\cite{zhang2018top}\cite{han2018advanced}and image question answering systems\cite{antol2015vqa}\cite{ren2015exploring}\cite{yang2016stacked}. Spatial transform network (STN) is an explicit processing module designed for networks that can handle transformations such as rotation invariance and scale invariance\cite{jaderberg2015spatial}. STN can be used in any standard neural network architecture to ensure that the model can acquire the ability of spatial transformation as a layer. There is no top-down attention mechanism in STN, since the ability of it to transform the target's space is based on the feedforward neural network. Conversely a better model is able to achieve  a whole target by use of the top-down attention mechanism to position and transform each individual in this target\cite{anderson2017bottom}\cite{stollenga2014deep}. Otherwise, it can cause image distortion during the spatial transformation and affect the following classification result\cite{Baluch2011Mechanisms}\cite{Zhang2015Efficient}. And with the number of geometric prediction layers increasing, it can cause undesired boundary effects\cite{Alimardani2007Three}\cite{Hu2015Multiple}.In this paper, we combine STN network with a Long Short-Term Memory (LSTM) network to create a LSTM-STN network that can resolve the problem of top-down attention mechanism. In the LSTM-STN structure, The LSTM network generates a series of time-step based outputs based on the input of the original images\cite{xingjian2015convolutional}\cite{graves2005framewise}\cite{gers2001lstm}, which are used as input data for the STN network.Then the STN outputs a series of transformed images,which are later used for classification in the subsequent CNN network layer. Recursion of steps in unit time step and transformation of previous time-step through LSTM network enable the LSTM-STN to sequentially catch the saliency elements. Since these significant element regions are smaller than the entire picture, the image is down-sampled with the LSTM-STN network to ensure that the resolution is essentially unchanged.
We organize the paper as follows:First of all, we give a review of multiple improved STN networks (for the top-down attention mechanism problem) in Section.2. Then we give an overview of Spatial Transformer Networks and describe our proposed LSTM-STN in detail in Section.3. In Section.4, we show the operation results of different kinds of images and compare these results with the output of RNN-STN method. Finally, we give our conclusion in Section.5.

\section{Related work}

Kuen et al.2016 proposed an recurrent attentional networks for saliency detection to detect the objects with a top-down mechanism\cite{kuen2016recurrent}.Chen-Hsuan Lin et al.2016 introduced the IC-STNs attention mechanism based on LK algorithm and Inverse Compositional Variant Method to eliminate distortion with less model capacity\cite{lin2016inverse}. Ba et al.2014 combined a nondifferentiable attention mechanism with a RNN and used it for classification.\cite{ba2014multiple} Anderson et al.2017 put forward bottom-up and top-down attention for image caption to amend CNN network and obtained improved top-down attention mechanism results\cite{anderson2017bottom}.

\begin{figure*}[t]
\centering
\includegraphics[width=17cm,height=5cm]{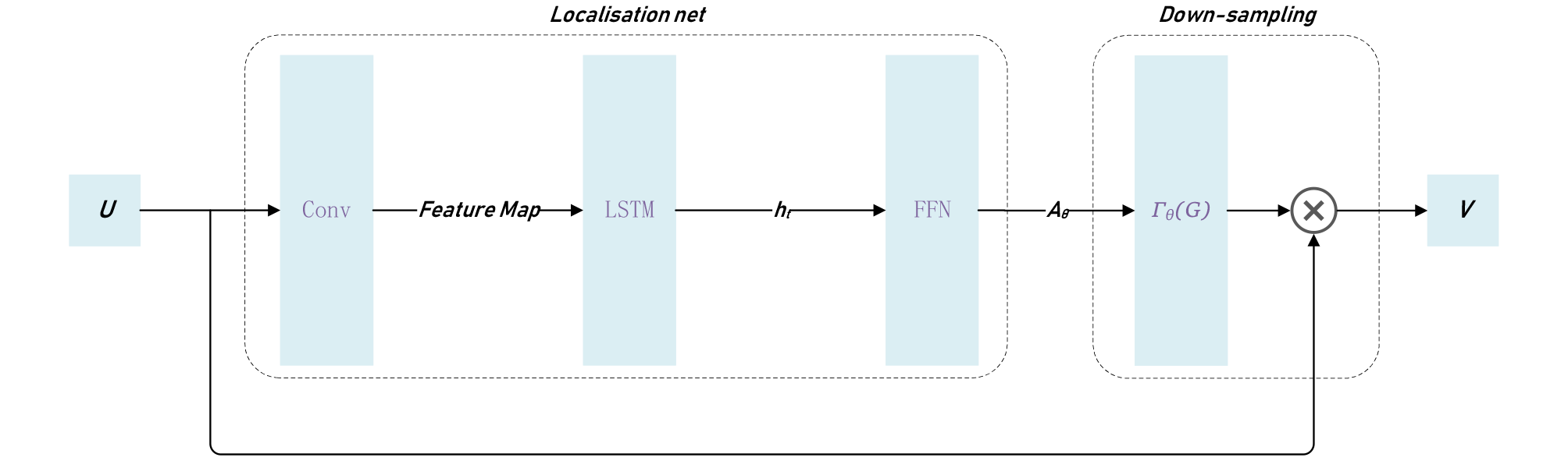}
\caption{The whole process of processing the image by the LSTM-STN model.$U$ is the input image, and $V$ are the transformed elements in a sequence.}
\end{figure*}

\section{Approach}
\subsection{Spatial Transformer Network}
STN consists of three main parts: localization network, parameterized sampling grid and differentiable image sampling. The role of the localization network is to generate a spatially transformed parameter $A_\theta$ through a subnetwork (fully-connected or convolutional network with a regression layer). The form of $A_\theta$ can be varied. For example, if we need to implement a 2D affine transformation (zoom, rotation and skew of the input), then $A_\theta$ would be the output of a 6-dimensional $(2\times 3)$ vector. $A_\theta$ can be described as follows:

\begin{center}
     \begin{equation}
    A_\theta=f_{loc}(U)=\left[\begin{matrix}
     \theta_{11}&\theta_{12}&\theta_{13}\\
     \theta_{21}&\theta_{22}&\theta_{23}\\
     \end{matrix}\right]
     \end{equation}
\end{center}

$U$ is the input of STN in the form of $[H\times W\times C]$ (height, width, channels), $A_\theta$ will make a format adjustment according to different affine transformation. 
Supposing the coordinates of each pixel in $U$ (not limited to the input original image, the feature map output by other layers included) are $\left(x_i^s,y_i^s\right)$ and that of each pixel in $V$ size $[H\times W\times C]$ are $\left(x_i^t,y_i^t\right)$. The spatial transformation function $\Gamma_\theta$ is an affine transformation function, the conversion relationship between $\left(x_i^s,y_i^s\right)$ and $\left(x_i^t,y_i^t\right)$ can be written as:

\begin{center}
    \begin{equation}
    \left(\begin{matrix}x_i^s\\y_i^s\end{matrix}\right)=\Gamma_\theta\left(G_i\right)=A_\theta\left(\begin{matrix}x_i^t\\y_i^t\end{matrix}\right)
    \end{equation}
\end{center}

After calculating $\Gamma_\theta$, we can get $V$ from $U$ by the following formula.

\begin{center}
    \begin{equation}
        \!V_i^c=\!\sum_n^H\!\sum_m^W U_{nm}^c max\left(0,\!1-\!\left|x_i^s-\!m\right|\!\right)
        max\left(0,\!1-\!\left|y_i^s-\!n\right|\!\right)
    \end{equation}
\end{center}

\begin{figure}[htb]
\centering
\includegraphics[width=8cm,height=10cm]{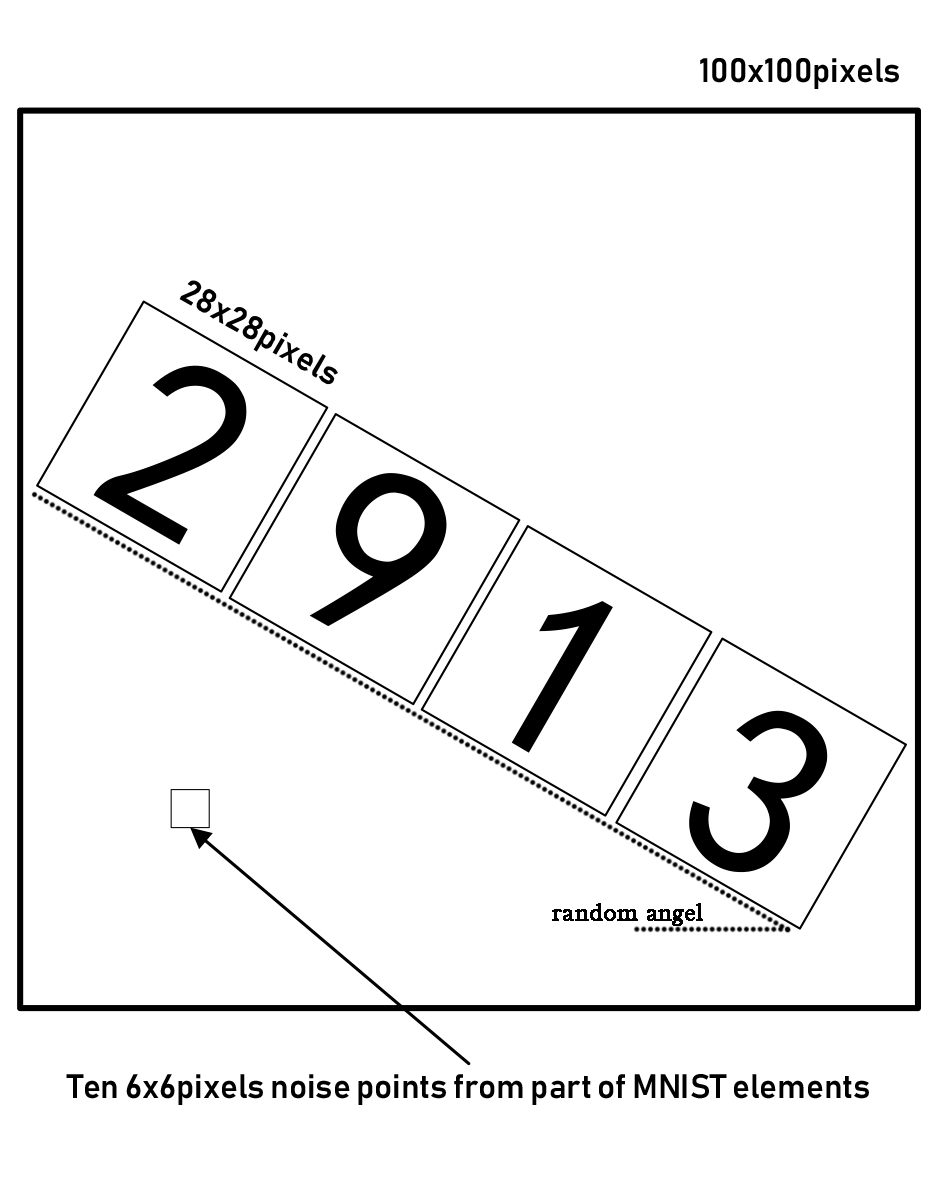}
\caption{The data set for this model test,which is composed of 4 random MNIST numbers according to a specified format.}
\end{figure}

\begin{figure*}[htb]
\centering
\includegraphics[width=17cm,height=9cm]{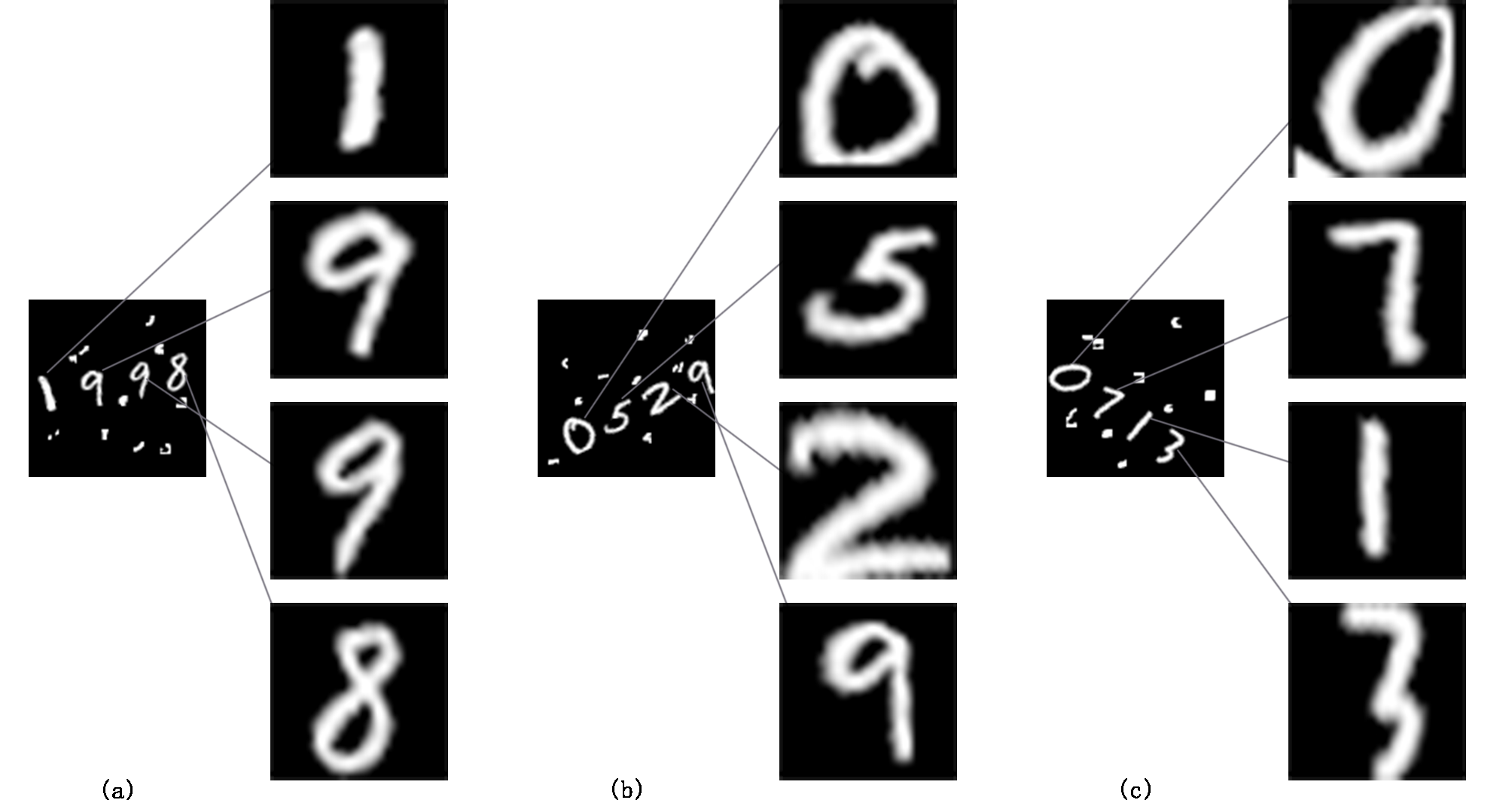}
\caption{Three sample images(a)(b)(c) processed by the LSTM-STN model with down-sampling factor 2. The transformed elements is used in the following CNNs}
\end{figure*}

After obtaining $V$, we need to derive $U$, $x_s$, $y_s$ to carry out the backward propagation of the network according to loss.

\subsection{LSTM-STN method}
Considering the lack of top-down attention mechanism in the classic STN method of feed-forward convolutional neural network, we modify the model by adding a LSTM prediction layer before the STN as follows: 1) get the feature map $f_{map}$ of $U$ through a convolutional network, 2) modify the localization network $f_{loc}$ by adding the LSTM method. 3)use ht as an input to calculate transformation parameter matrix $A_\theta$ in FFN.

\begin{center}
    \begin{equation}
        f_{map}=f_{conv}\left(U\right)
    \end{equation}
    \begin{equation}
        \left(c_t,h_t\right)=f_{LSTM}\left(f_{map},C_{t-1},h_{t-1}\right)
    \end{equation}
    \begin{equation}
        A_\theta=g\left(h_t\right)
    \end{equation}
\end{center}
The structure of the above improved STN can be illustrated by the "localisaton net" part of Fig1.

\subsection{Down-sampling}
For an image $U$ in size of $H\times W$, it is d-sampled down to obtain a resolution image of $\left(\frac{H}{d}\right)\left(\frac{W}{d}\right)$ in size. And d should be the common divisor of H and W. If $U$ is in 2D matrix form, then the image pixels in the original $d\times d$ window will be turned into one pixel. The value of this pixel is the average of all the pixels in the window\cite{zhang2011interpolation}\cite{nguyen2015downsampling}. The number of sampled points from $U$ is:

\begin{center}
    \begin{equation}
    n_{points}=\left(\frac{H}{d}\right) \left(\frac{W}{d}\right)    
    \end{equation}
\end{center}

The down-sampling is done before the input of classification network to ensure the correct rate. The whole process of processing the image by the LSTM-STN model is shown in Fig 1.
\section{Evaluation}
Learning-based approach requires lots of training samples to generalize new examples well. We train our model on the MNIST’s training data-set containing 60000 traning samples and test the model effects in test dataset MNIST’s testing data-set involving 10000 test samples. Based on the elements in the MNIST dataset, we created a data-set for this model test. On each of the independent canvases (size $100\times 100$ pixels), we selected 4 random numbers in MNIST(size $28\times 28$ pixels), the first of which is placed on the canvas at random rotation angles and positions. The following numbers are placed at the same rotation angle to ensure that they do not overlap, and that the final number field does not exceed the canvas size. Finally, we place 10 noise handwriting image blocks (each size is $6\times 6$ pixels) on the canvas. We create 70000 examples for training, 20000 for validation, and 2000 for testing. Fig2 shows the format of the data set for this model.Fig3 shows two examples of the spatial transformation process. 

In order to test the merits of the operation results, we train the traditional backward propagation spatial transformation network as a contrast model for testing. The entire classification network is divided into four layers, namely: convolution, maximum pooling, dropout, and a fully connected layer with 400 units.Eventually the results are output by a series of separate softmax. And the convolutional layer has 96 filters of size $[3\times 3]$ in total.
The LSTM-STN uses a Long Short-Term Memory (LSTM) with 256 cells. The time step of LSTM is 4, and in each step the LSTM cell uses $C_{t-1}$ and $h_{t-1}$ as input to deliver the output $h_t$ through a linear layer to spatial transform layer. The LSTM-STN is followed by a CNN with 32 filters used as classification layer. In the LSTM-STN model, the localization network contains 4 max-pool convolutional layers, each with 20 filters of size $[3\times3]$.

\begin{center}
    \begin{table}[h]
    \centering
    \caption{Error rate of different models when identifying data in data sets, d is the down-sampling parameter}
    \begin{tabular}{cc}
    \toprule  
    \multicolumn{2}{c}{\emph{Mnist Sequence Example}}\\
    Model&Error rate(\%)\\
    \midrule  
    LSTM-STN-CNN d=1& 1.7\\
    LSTM-STN-CNN d=2& 1.6\\
    LSTM-STN-CNN d=3& 1.8\\
    LSTM-STN-CNN d=4& 2.2\\
    FFN-STN-CNN d=1& 3.7\\
    FFN-STN-CNN d=2& 2.2\\
    FFN-STN-CNN d=3& 2.9\\
    FFN-STN-CNN d=4& 5.6\\
    CNN&2.7\\
    \bottomrule 
    \end{tabular}
    \end{table}
\end{center}

\section{Conclusion}
By combining STN with LSTM,we can obtain a better result compared with the original STN model in the classification of objects. The LSTM-STN model can focus on each detected element in a sequence, and this model shows the classification ability when dealing with individual elements with some interference noise in the sequences, which can not be done in the original STN model. Also noteworthy is that the speed of training LSTM-STN model is close to original STN model because of the low capacity of LSTM layer. In the long run, our model is expected to play a role in detecting multiple overlapping objects in an image or video in the future work. While this suggests several directions for future research, the immediate benefits of our approach may be captured by reducing the impact of noise on the objects in order to be classified. Our approach can also implement an undistorted spatial transformation when multiple individual elements are included in sequences.

\bibliographystyle{IEEEtran}
\bibliography{mylib.bbl}

\end{document}